\documentclass[10pt,letterpaper,twocolumn]{article} %% two column, final layout

\usepackage{ol2}
\usepackage[draft]{hyperref}
\usepackage{amsmath}

\begin{document}

\twocolumn[ %% activate for two-column option

\title{Negative refraction and spatial echo in optical waveguide arrays}

\author{Ramaz Khomeriki$^{1,2,*}$ and Lasha Tkeshelashvili$^{2,3}$}

\address{
$^1$Max-Planck Institute for the Physics of Complex Systems, \\ N\"othnitzer Str. 38, 01187 Dresden, Germany \\
$^2$Physics Department, Tbilisi State University, \\ Chavchavadze 3, 0128 Tbilisi, Georgia \\
$^3$Institute of Physics, Tbilisi State University, \\ Tamarashvili 12, 0169 Tbilisi, Georgia \\
$^*$Corresponding author: khomeriki@hotmail.com
}

\begin{abstract}
The special symmetry properties of the discrete nonlinear
Schr\"odinger equation allow a complete revival of the initial
wavefunction. That is employed in the context of stationary
propagation of light in a waveguide array. As an inverting system
we propose a short array of almost isolated waveguides which cause
a relative $\pi$ phase shift in the neighboring waveguides. By
means of numerical simulations of the model equations we
demonstrate a novel mechanism for the negative refraction of
spatial solitons.
\end{abstract}

\ocis{190.4370, 190.6135.}

%%190.4370 Nonlinear optics, fibers
%%190.6135 Spatial solitons

 ] %% activate for two-column option

\noindent
The optical properties of nano-structured metallic systems, often
called metamaterials, can be tailored to almost any
need~\cite{busch2007}. The left-handed materials with negative
refractive index represent a striking example of such kind of
artificial structures~\cite{pendry2004}. It is important that the
negative refraction is predicted to exist for nonlinear
metamaterials as well~\cite{agranovich2004}. Apparently, the
concept of negative refraction implies great potential for the
control and manipulation of optical solitons~\cite{busch2007,lederer2008}.
It should be noted, however, that according to the Kramers-Kronig relations,
the left-handed materials are inherently absorptive at the relevant
frequencies~\cite{stockman2007}. Although the dielectric nano-structures cannot be
described by effective medium parameters, it is still possible to apply the generalized
Snell's refraction law to such systems (e.g. see \cite{szameit2008a}).
In this sense, the negative refraction can be realized in periodic dielectric
systems as well~\cite{busch2007}. The physical mechanisms that govern the
electromagnetic fields in dielectric nanostructures are different
from those in the metallic systems, and therefore, the constraint
based on the Kramers-Kronig relations does not apply. 

Recently, in \cite{rosberg2005} the negative refraction of optical beams 
at interfaces between bulk dielectrics and tilted photonic lattices was 
studied both theoretically and experimentally. 
Moreover, it was shown that the discussed effect persists in the presence 
of nonlinear self-focusing. The negative refraction was predicted for linear 
waves at interfaces between two photonic crystal waveguide arrays 
as well \cite{locatelly2006}. Later, in \cite{prilepsky2011} it was demonstrated 
that the solitary wave refraction can be effectively controlled in nonlinear optical
lattices. Here, we show that the negative refraction and spatial echo for solitons 
can be realized in nonlinear periodic dielectrics.
In particular, the soliton propagation in an array of coupled
optical waveguides~\cite{christodoulides2003,lederer2008} is
chosen as a model system. We demonstrate that the properly
designed scatterer in the system causes the desired relative phase
shift in the neighboring waveguides, and leads to the negative
refraction of solitons in the nonlinear optical lattices.

Our consideration is based on the fact that in quantum mechanics
the Schr\"odinger equation is invariant under
the simultaneous transformations $t\rightarrow -t$ and
$\psi\rightarrow\psi^*$. Therefore, the conjugation of the system
wave function at the arbitrary time moment leads to the reversal
of the system dynamics. Alternatively, it is possible to change
the sign of the whole Hamiltonian. For example, this can be done
in the context of cold atoms in optical lattices by simultaneously
changing the signs of hopping and interaction constants
\cite{ch}. The formal analogy of the quantum-mechanical
systems with light propagation in waveguide arrays
\cite{longhi2009} allows one to expect the similar effects in the
context of optics. In particular, the complete image reconstruction
in a system of coupled linear waveguides was demonstrated
theoretically \cite{longhi2008} and realized experimentally \cite{szameit2008b}.

In order to explore the above mentioned analogy, let us start with the
Discrete Nonlinear Schr\"odinger (DNLS) equation:
\begin{equation}
i\frac{\partial\psi_m}{\partial
z}=J\left(\psi_{m+1}+\psi_{m-1}\right)+\theta(z)|\psi_m|^2\psi_m,
\label{sch}
\end{equation}
where $\psi_m$ stands for an on-site function of its argument, and
$\theta(z)$ is a stepwise function defined as $\theta(z<0)=-1$ and
$\theta(z>0)=1$. It is easy to see that this equation is invariant
under the simultaneous transformations $\psi_m\rightarrow
(-1)^m\psi_m$ and $z\rightarrow-z$. Furthermore, let us suppose
that Eq. \eqref{sch} is valid in the interval $|z|>\ell$ and allow
the system to evaluate from the initial state $\psi_m(-L)$ to
$\psi_m(-\ell)$. Now, if the system can somehow be brought to the
state $\psi_m(\ell)=(-1)^m\psi_m(-\ell)$, the initial state
re-emerges at $z=L$ with $\psi_m(L)=(-1)^m\psi_m(-L)$. The
transition from $\psi_m(-\ell)$ to
$\psi_m(\ell)=(-1)^m\psi_m(-\ell)$ can be achieved by evaluating
the system according to the following equation:
\begin{equation}
i\frac{\partial\psi_m}{\partial
z}=\left[(-1)^m+\frac{1}{2}\right]\frac{\pi}{2\ell}\psi_m,
\label{simp}
\end{equation}
in the interval $|z|<\ell$. It should be noted that the neighboring sites gain
$\pi$ phase difference in this interval.

Therefore, for our purpose, the waveguide array design must be governed by
\eqref{sch} in the interval $|z|>\ell$, while for $|z|<\ell$ the approximate
model equation \eqref{simp} must be valid. The discrete equations
\eqref{sch} and \eqref{simp} represent good approximations for the cold atom
dynamics in deep harmonic potential or light propagation in
optical waveguide arrays. A more general theoretical framework is given by:
\begin{equation}
i\frac{\partial\Psi}{\partial z}=\frac{\partial^2\Psi}{\partial
x^2}+\delta n\Psi+\chi(z)|\Psi|^2\Psi, 
\label{GP1}
\end{equation}
for $|z|>\ell$. Here, $ \delta n=W\cos (x)$, and $\chi(z)=-\chi(-z)$ stands for
the Kerr-type nonlinearity constant. Moreover, for $|z|<\ell$:
\begin{equation}
i\frac{\partial\Psi}{\partial z}=\frac{\partial^2\Psi}{\partial
x^2}+\delta n\Psi,
\label{GP2}
\end{equation}
where $\delta n = W'\cos(x)+(\pi/4\ell)\cos (x/2)$. 
In these equations the variables $x$ and $z$ are scaled in units of $1/K$
and $k/K^2$, respectively. $K$ represents the inverse spacing of
harmonic transverse modulations of the refractive index, and $k$
is the carrier wavenumber given by $k=n_0\omega/c$. Here, $\omega$
is the laser beam frequency, and $n_0$ is the averaged refractive
index. Note that, $\delta n \equiv k^2\left[n(x)-n_0\right]/n_0K^2$
defines the scaled relative variation of the linear refractive index in the 
corresponding regions of the waveguide array.

\begin{figure}[t]
\centerline{\includegraphics[width=7.5cm]{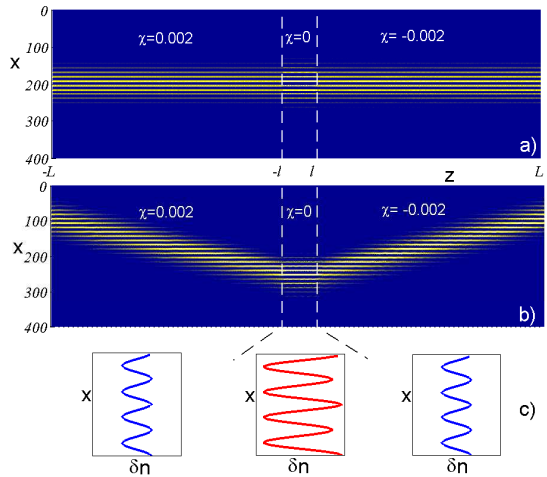}}
\caption{Light intensity distributions for a) and b) represent the results of numerical
simulations with different boundary conditions of \eqref{GP1} for
$L<|z|<\ell$ and of \eqref{GP2} for $|z|<\ell$. Here, $L=750$ and
$\ell=50$. The values of nonlinear coefficients are shown in the
graphs. Dashed vertical lines indicate the boundaries of the
scatterer. c) schematically displays the refractive index profiles
in the corresponding arrays (see the text for the values of
modulations).}
\label{negative1}
\end{figure}

Let us first examine Eq.~\eqref{GP1}. Note that, as far as the case of deep
lattices is concerned, the wavefunction can be assumed to be localized
around the maximums of refractive index profile, i.e. at the points
$x=2\pi m$ ($m$ is integer). Moreover, the fundamental mode for the $m$th potential
well can be approximated by:
\begin{equation}
\varphi_m(x)\sim \exp\left\{-\sqrt{W}(x -2\pi
m)^2/\sqrt{8}\right\}. 
\label{7}
\end{equation}
In this case the tight-binding approximation applies
\cite{smerzitight,yuri,mark1,mark2}, and the full wave function $\Psi(x,z)$ can be
written as follows:
\begin{equation}
\Psi(x,z)=\sum_m\psi_m(z) \varphi_m(x). 
\label{6}
\end{equation}
Furthermore, assuming that the mode overlap of the
neighboring sites is small, \eqref{GP1} reduces to the DNLS equation
\eqref{sch} where the coupling constant $J$ can be easily determined
from the following expression \cite{smerzitight}:
\begin{equation}
J=\int\left[\delta
n(x)\varphi_m\varphi_{m+1}-\frac{\partial\varphi_m}{\partial
x}\frac{\partial\varphi_{m+1}}{\partial x} \right]d x. 
\label{9}
\end{equation}
Here the normalization of eigenmodes $\varphi_m$ in \eqref{7} is
chosen such that in DNLS equation (see Eq.~\eqref{sch}) the
nonlinear coefficient value becomes exactly one.

In the case of Eq. \eqref{GP2} the lattice of depth $W'$ is
modulated with the amplitude $\pi/4\ell$. As far as this
modulation is assumed to be small, similar to Eq.~\eqref{7} we can
write $\varphi_m(x)\sim \exp\left\{-\sqrt{W'}(x -2\pi
m)^2/\sqrt{8}\right\}$. Nevertheless, the harmonic modulation
affects the corresponding eigenenergies of the modes and leads to
a staggered energy structure. Using again the expansion \eqref{6}
for large values of $W'$ we obtain the negligible overlap between
neighboring modes $J\rightarrow 0$. Therefore, we can equate the
hopping terms to zero which finally gives the discrete
equation~\eqref{simp}.

We performed the numerical simulations based on Eqs.~\eqref{GP1} and
\eqref{GP2} with the initial conditions at $z = -L$ ($L\gg\ell$).
Initially, the launched spatial soliton propagates along the trajectory $z = vx$.
Here, the parameter $v$ determines the direction of the optical beam.
Due to the effective index mismatch, the scattering process takes place 
at the interfaces of the waveguide array region $-\ell<z<\ell$.
That causes the change of the beam propagation trajectory to $z=-vx$, and
eventually, results in the negative refraction of the soliton. Moreover, 
the arbitrary initial profile of the wave packet at $z=-L$ is expected 
to be reproduced at $z=L$.

In the numerical simulations (see Fig.~\ref{negative1}) we choose $L=750$ and
$\ell=50$. In the region $\ell<|z|<L$ the refractive index modulation is
$\delta n=\cos(x)$, and the nonlinear coefficient is $|\chi|= 0.002$
(for $-L<z<-\ell$ we have $\chi<0$, while $\chi>0$ for $\ell<z<L$).
Moreover, in the range $|z|<\ell$ refractive index is modulated according to
$\delta n=3\cos(x)+(\pi/4\ell)\cos(x/2)$, and the nonlinear
coefficient $\chi=0$. The phase modulation of the initial wave function
determines the spatial soliton propagation "velocity". As expected,
Fig.~\ref{negative1} (a) shows that the soliton propagation is not affected by the
the thin scatterer if the incident angle is normal ($v=0$). Furthermore,
Fig.~\ref{negative1} (b) demonstrates that the soliton with $v\not=0$ experiences
the negative refraction after the interaction with the scatterer.

\begin{figure}[t]
\centerline{\includegraphics[width=7.5cm]{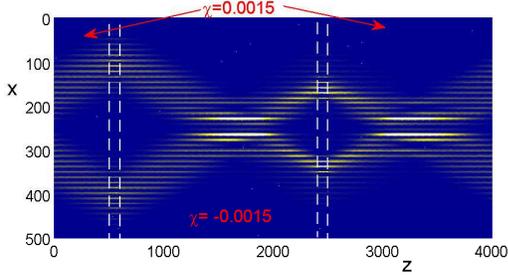}}
\caption{The same as in
Fig.~\ref{negative1} but with two solitons and two scatterers.
Note that the nonlinearity is negative in the region between the
scatterers while it is positive on the left and right sides.}
\label{negative3}
\end{figure}

Nevertheless, although the output intensity distribution is the
same among waveguides as at the input, the phase distribution is
modified. In particular, as it was stressed above,
$\psi_m(L)=(-1)^m\psi_m(-L)$. In order to have the exact revival
of the wavefunction the second scatterer must be introduced. Then,
if the second scatterer is identical to the first one, the soliton
"lensing" can be realized (see Fig.~\ref{negative3}).

It should be stressed that the presented analysis of the refraction effects
in $1+1$ dimensions can be extended to the case of $1+2$ dimensions as well.
In particular, if the propagation direction coincides with the $z$-axis,
the scatterer in $1+2$ dimensions represents a thin film in $xy$ plane.
The refractive index profile is given by:
\begin{equation} \delta
n=W'\cos(x)\cos(y)+\frac{\pi}{4\ell}\cos(x/2)\cos(y/2),
\label{GP3}
\end{equation}
and the thickness of the film is $2\ell$.

In summary, the presented results of numerical simulations clearly
demonstrate that the negative refraction of spatial solitons can
be realized in an array of coupled optical waveguides. This effect
is related to more general phenomenon which can be interpreted as
spatial echo. Based on the analysis of the theoretical model we
suggest the inverting devise that allows the complete revival of
the initial wave packet. It must be stressed that the discussed
effects may have direct implications for the telecommunication
applications. Similar effects can also be realized in different
physical systems which can be described by the discrete nonlinear Schr\"odinger
equation. These include Bose-Einstein condensates in deep optical
lattices, coupled defects in photonic crystals, etc.

This work is supported by joint grant No 09/08 from RNSF (Georgia)
and CNRS (France); The funding from  Science and Technology Center
in Ukraine (Grant No 5053) is also acknowledged.

\end{document}